\documentclass[%
 reprint,superscriptaddress,
 amsmath,amssymb,prx]{revtex4-2}

\usepackage{physics}
\usepackage{xcolor}
\usepackage{soul}
\usepackage{xtab,afterpage,longtable}
\usepackage{lipsum}
\usepackage{graphicx}
\usepackage{dcolumn}
\usepackage{booktabs} 
\usepackage{multirow}
\usepackage{bbm}  
\usepackage{color}
\usepackage{csquotes}
\usepackage{hyperref}
\usepackage{soul}

\makeatletter
\renewcommand*{\@fnsymbol}[1]{\ensuremath{\ifcase#1\or \dagger\or *\or \ddagger\or \mathsection\or \mathparagraph\or \|\or **\or \dagger\dagger\or \ddagger\ddagger\else\@ctrerr\fi}}
\makeatother

\begin{document}

\title{Thermal Equilibrium Vacancy Concentration in an Alloy with Chemical Short-Range Order}

\author{Hao Tang}
\thanks{These authors contributed equally to this work.}
\affiliation{Department of Materials Science and Engineering, Massachusetts Institute of Technology, MA 02139, USA}

\author{Hoje Chun}
\thanks{These authors contributed equally to this work.}
\affiliation{Department of Materials Science and Engineering, Massachusetts Institute of Technology, MA 02139, USA}

\author{Rafael G\'{o}mez-Bombarelli}
\affiliation{Department of Materials Science and Engineering, Massachusetts Institute of Technology, MA 02139, USA}

\author{Yuri Mishin}
\thanks{Corresponding authors: liju@mit.edu, ymishin@gmu.edu}
\affiliation{Department of Physics and Astronomy, MSN 3F3, George Mason University, Fairfax, VA 22030, USA}

\author{Ju Li}
\thanks{Corresponding authors: liju@mit.edu, ymishin@gmu.edu}
\affiliation{Department of Materials Science and Engineering, Massachusetts Institute of Technology, MA 02139, USA}
\affiliation{Department of Nuclear Science and Engineering, Massachusetts Institute of Technology, Cambridge, MA 02139, USA}

\date{\today}

\begin{abstract}
The equilibrium vacancy concentration in multi-principal element alloys remains a controversial and nontrivial subject, primarily because of chemical complexity and chemical short-range order (CSRO). 
 Here we derive an exact expression that is amenable to atomistic calculations, using multiple perspectives.
We applied this expression to equiatomic CrCoNi alloys in the face-centered cubic structure. The derived equilibrium vacancy concentration is used in our recent work~\cite{chun2024learning}, which predicts the chemical short-range order formation timescale consistent with experimental observation. 
The results demonstrate the practical utility of the approach for predicting equilibrium vacancy concentrations in compositionally complex alloys.
\end{abstract}

\maketitle

\section{Introduction}
Point defects in crystalline materials usually play an important role in various physical properties, such as mass transport, phase stability, electric conductivity, and mechanical response~\cite{freysoldt2014first, peng2022vacancy}. The equilibrium concentration of vacancies is a central thermodynamic quantity that determines the intrinsic rate scale for atomic diffusion and underpins kinetic theories across metals, ceramics, and semiconductors~\cite{van2010vacancy}. Although vacancy thermodynamics in elemental and binary crystals is well studied~\cite{behara2024role, wang2017thermodynamics,belak2015effect}, modern alloy design increasingly targets multicomponent systems, where chemical complexity gives rise to spatially heterogeneous local environments, nontrivial bonding energetics, and chemical short-range order (CSRO)~\cite{george2019high}.

The precise formulation to compute the thermal equilibrium vacancy concentration $c_{\rm V}^{\rm eq}=X_{\rm V}^{\rm eq}/v$, where $v$ is the mean atomic volume (molar  average  volume) of a multi-element alloy, has been controversial~\cite{zhang2021statistical, zhang2022ab}. The difficulty in this problem  is that in multi-element alloy, the vacancy will be dressed by the CSRO, so the vacancy formation energies also have configurational multiplicity. 
Recent study shows that the vacancy formation energies in multi-principal element alloys strongly depend on the local chemical environment, leading to CSRO-dependent equilibrium vacancy concentrations~\cite{li2024vacancy}. Therefore, recent efforts have been focusing on building practical frameworks for computing vacancy formation energetics in compositionally complex alloys using statistical mechanics methods~\cite{lee2025modeling,li2024vacancy}. 

In this work, we develop semiclassical partition function based approach for arbitrary number of chemical elements, with and without vacancies, and express quantities such as the vacancy formation volume, formation enthalpy and formation entropy. We then consider numerical approaches to computing $X_{\rm V}^{\rm eq}(T)$, when we use a small, periodic boundary condition (PBC) supercell with $\tilde{n}$ sites ($\tilde{n}$ takes typical values like 256, 500, 864, 1372, 2048, ... for FCC single crystal) at zero pressure to do the calculations. To derive $X_{\rm V}^{\rm eq}$, we run a gedankenexperiment of grand canonical Monte Carlo (GCMC) simulation~\cite{tanguy2011monte}, where a much larger system of $M^3\tilde{n}$ sites can exchange atoms with external chemical potential reservoir $\{\mu_c\}, c=1..C$, to create/backfill vacancies.  We show that the mono-vacancy equilibrium fraction can be well approximated by $X_{\rm V}^{\rm eq}(T)\approx \left\langle \sum_{i=1}^{\tilde{n}} e^{-\Delta f({\rm site}\; i)/k_{\rm B}T}/\tilde{n} \right\rangle$, where $\Delta f({\rm site}\; i)$ is the local free-energy difference before/after removing a site-$i$ atom of chemical type $c({\rm site}\; i)$ in an initially fully dense PBC supercell, plus $\mu_{c({\rm site}\; i)}$ of external reservoir, as long as (a) the spectrum of $\Delta f({\rm site}\; i)$ has a non-negative support, which should be satisfied under most circumstances, and (b) the equilibrium vacancy concentration is dilute: $X_{\rm V}^{\rm eq}\ll 1$, and divacancy (V$_2$), trivacancy (V$_3$), etc. concentrations are even much more so, $X_{\rm V_2}^{\rm eq}, X_{\rm V_3}^{\rm eq}\ll X_{\rm V}^{\rm eq}$. The $\left\langle \cdot \right\rangle$ averaging must be generated based on plenty of prior atom-preserving exchanges in the $\tilde{n}$-site supercell, i.e. sampling of a well-equilibrated, fully dense crystal, where CSRO between the real atoms are fully developed. Justifications of the Metropolis GCMC gedankenexperiment are provided. Finally, we analyze the consequence of the spectrality (dispersion) of the site-specific vacancy formation free energy 
$\Delta f({\rm site}\; i)$, and show that {\em vacancy-affiliated} CSRO is an important feature for the numerical value of $X_{\rm V}^{\rm eq}(T)$.  The numerical example of  equiatomic CrCoNi alloy in FCC structure is provided, at various temperatures.

\section{Conceptual framework and Derivations}
\label{conceptual_framework}

We first consider an alloy phase with $c=1..C$ real element types.  For example $C=3$ and $c$=1 means Cr, $c$=2 means Co, and $c$=3 means Ni. This alloy phase has only 1 sublattice, say FCC. Initially let us not consider lattice vacancies.  The statistical mechanics of such lattice is well established.  At external pressure $P=0$, the Helmholtz free energy $F$ is equal to the Gibbs free energy $G$:
\begin{equation}
  F\equiv -k_{\rm B}T\ln Z = G = \sum_{c=1}^C N_c \mu_c
 \label{FandGidentity}
\end{equation}
where $Z$ is the partition function:
\begin{equation}
    Z\equiv \int \frac{dp^N dq^N}{h^{3N}\prod_{c=1}^C N_c!} \exp(-\mathcal{H}/k_{\rm B}T)
    \label{Definition1Z}
\end{equation}
and we took the semiclassical expression where $\mathcal{H}$ is the Hamiltonian, $h$ is the Planck constant and $\prod_{c=1}^C N_c!$ takes into account the indistinguishability of particles of the same type (ignoring isotopes), and $p^N, q^N$ are the conjugate momenta and position, respectively, with 
\begin{equation}
    N \equiv \sum_{c=1}^C N_c
\end{equation}

Given 
\begin{equation}
    \mathcal{H} = \sum_{c=1}^C \frac{p_c^2}{2m_c} + U(q^N)
\end{equation}
where $U(q^N)$ is the potential energy, 
we can integrate out (\ref{Definition1Z}) and obtain
\begin{equation}
    Z= \prod_{c=1}^C (2\pi m_c k_{\rm B} T/h^2)^{3N_c/2} \int \frac{dq^N}{\prod_{c=1}^C N_c!} \exp(-U/k_{\rm B}T)
    \label{Definition2Z}
\end{equation}
The prefactor can be identified as the thermal wavelength of type-$c$ atom, $\lambda_{Tc}$:
\begin{equation}
    \lambda_{Tc}\equiv  \frac{h}{\sqrt{2\pi m_c k_{\rm B} T}}
    \label{ThermalWavelength}
\end{equation}
\begin{equation}
    Z= \prod_{c=1}^C \left( \frac{1}{\lambda_{Tc}^{3}} \right)^{N_c} \int \frac{dq^N}{\prod_{c=1}^C N_c!} \exp(-U/k_{\rm B}T)
    \label{Definition3Z}
\end{equation}
Combining (\ref{FandGidentity}), we get fundamental identity
\begin{equation}
    1 = \prod_{c=1}^C \left( \frac{1}{\lambda_{Tc}^{3}} \right)^{N_c} \int \frac{dq^N}{\prod_{c=1}^C N_c!} 
    e^{-\frac{U(q^N)-\sum_{c=1}^C N_c \mu_c}{k_{\rm B}T}}
    \label{FundamentalIdentity}
\end{equation}
at thermal equilibrium. The interpretation is that $e^{-\frac{U(q^N)-\sum_{c=1}^C N_c \mu_c}{k_{\rm B}T}}$ can be regarded as the {\rm probability density} (non-negative and normalized) of microstates at thermal equilibrium.  By generalizing from canonical to grand canonical ensemble (and performing so-called saddle integration over thermal fluctuations), $e^{-\frac{U(q^N)-\sum_{c=1}^C N_c \mu_c}{k_{\rm B}T}}$ can also be interpreted as the probability density in the grand canonical ensemble, where there is an external source/sink of real atoms, and by trading with it, one can freely remove/add/exchange atoms of any type from the system of interest.

Based on this interpretation, one can carry out Monte Carlo (MC) where $\{N_c\}$ are fixed, or Grand Canonical Monte Carlo (GCMC)\cite{MoeiniArdakaniTL21} where $\{N_c\}$ are variable, in atomistic simulation driven by an interatomic potential $U(q^N)$.\cite{TakamotoOLL23}  Note that for a crystalline solid, the $P=0$ condition means that different-$\{N_c\}$ configurations, and even the same-$\{N_c\}$ configurations but trapped in different local crystalline site arrangements, must have different volume $V(q^N)$ at thermal equilibrium.  How $V\equiv \langle V(q^N) \rangle$ depends on $\{N_c\}$ can in fact be regarded as the partial molar volume 
\begin{equation}
    v_c \equiv \left. \frac{\partial V}{\partial N_c} \right|_{\{N_{c'\neq c}\},T,P=0}
\end{equation}
or the Vegard strain.  The mean atomic volume (molar average volume) of this multi-element alloy is just 
\begin{equation}
    v = \sum_{c=1}^C X_c v_c = \sum_{c=1}^C \frac{N_c}{N} v_c,
\end{equation}
due to the well-known relation between partial molar and molar average quantities.  The relationship between concentration (unit $\#$/m$^{-3}$) and fraction (dimensionless) of any species is simply a division by $v$.

To separate out the perturbational atomic displacements from the site arrangement degrees of freedom in a crystalline lattice, we can write
\begin{equation}
 q^N \equiv s^N + \delta q^N
 \label{ConfigurationalVibrational}
\end{equation}
We can do constrained integration over $\delta q^N$, fixing the site arrangement degrees of freedom, as
\begin{equation}
    1 = \sum_{s^N} e^{-\frac{f(s^N)}{k_{\rm B}T}}
    \label{FundamentalIdentity2}
\end{equation}
where $f(s^N)$ is the vibration-integrated constrained free energy:
\begin{equation}
    f(s^N) \equiv -k_{\rm B}T\ln\left( \int \frac{d(\delta q^N)}{\prod_{c=1}^C \lambda_{Tc}^{3N_c}} 
    e^{-\frac{U(s^N + \delta q^N)-\sum_{c=1}^C N_c \mu_c}{k_{\rm B}T}} \right)
    \label{MicroscopicFreeEnergy}
\end{equation}
where the permutation symmetry factor $\prod_{c=1}^C N_c!$ is removed since trivially permuted labeled site arrangement is regarded as the same $s^N$.  Note that we use lower-case $f(s^N)$ instead of capital $F(s^N)$ because $s^N$ is still a {\em microstate}, albeit vibration-integrated.

In MC/GCMC\cite{MoeiniArdakaniTL21}, we need to have a good numerical approximant to (\ref{MicroscopicFreeEnergy}), and then use the microscopic free energy difference
\begin{equation}
  \Delta f \equiv  f(s^N_{\rm attempt}) - f(s^N)
  \label{MicroscopicFreeEnergDifference}
\end{equation}
to accept/reject an attempted new configuration $s^N_{\rm attempt}$, based on the Metropolis sampling of transitions $s^N\rightarrow s^N_{\rm attempt}$, which is derived based on detailed balance principle.  We note that chemical short-range order (CSRO) in the 
single-sublattice alloy will 
naturally arise out of such MC procedures.  CSRO is the lattice-equivalent of pair distribution function (PDF) in fluids\cite{LamLBFL21,LamLMFBL21}, and can be measured experimentally by diffraction methods.

Now we are ready to discuss about vacancies.  In a single-sublattice alloy, we can have a lattice site occupied by an atomic number $A=Z=0$ ``element", which we can count as $N_0$.  Generally speaking, $N_0\ll N_1, N_2, N_3$, the major elements.  Let us use $\sim$ for vacated system:
\begin{equation}
    \tilde{N} \equiv \sum_{c=0}^C N_c = N_0 + N
\end{equation}
We recognize that 
\begin{equation}
    \tilde{N} = N_{\rm site} \ge N
\end{equation}
where $N_{\rm site}$ is the number of atomic sites occupied by the solid, and it is greater than the number of real atoms in the system, meaning the solid ``looks and feels" bigger than the fully dense version, due to the localized ``porosities" inside. (As a side remark, strain and mechanical deformation cares more about $N_{\rm site}$ and how these sites are distributed spatially, not $N$, since our hands and eyes cannot make out individual atoms/vacancies or their chemical types).

It turns out that $\tilde{N} > N$ achieves lower $F$ than $\tilde{N} = N$ at finite $T$.  This phenomenon is called {\em equilibrium vacancy fraction}:
\begin{equation}
   X_{\rm V}^{\rm eq} \equiv \frac{N_0^{\rm eq}}{\tilde{N}}
\end{equation}
Let us develop a theory for $X_{\rm V}^{\rm eq}(T)$. Consider the same \begin{equation}
  G = F(\tilde{N}) = F(N_0; N_1, ..., N_C) \equiv -k_{\rm B}T\ln Z,
 \label{FandGidentityVacancy}
\end{equation}
pretty much exactly the same expressions as before (same equations and explanations from Eq. (\ref{Definition1Z}) to \ref{MicroscopicFreeEnergDifference}), but with added parametric freedom of $N_0$.  That is to say, atoms can occupy more sites.  We will use the notation 
\begin{equation}
  s^N \rightarrow s^{\tilde{N}}
\end{equation}
to denote a configuration $(N_0; N_1, ..., N_C)$ with porosity. The {\em value} of $F(N_0; N_1, ..., N_C)$ involves $dq^N$ integral, but its arguments are $\tilde{N}$.

Let us first imagine fixed $N_1, ..., N_C$. All the expressions stay the same, with an additional
\begin{equation}
    v_0 \equiv \left. \frac{\partial V}{\partial N_0} \right|_{\{N_{c\neq 0}\},T,P=0}
    \label{VacancyFormationVolume}
\end{equation}
$v_0$ is recognized as the vacancy formation volume (relaxed).  Similarly, the vacancy formation enthalpy  
\begin{equation}
    h_0 \equiv \left. \frac{\partial H}{\partial N_0} \right|_{\{N_{c\neq 0}\},T,P=0}
    \label{VacancyFormationEnthalpy}
\end{equation}
with $H=E$ at $P=0$.  Given $F=H-TS$, we can also define vacancy entropy as 
\begin{equation}
    s_0 \equiv \left. \frac{\partial S}{\partial N_0} \right|_{\{N_{c\neq 0}\},T,P=0}.
    \label{VacancyEntropy}
\end{equation}
It is well recognized that $\{ v_0, h_0, s_0, ... \}$ are partial molar quantities for the {\em defect}, on the same footing as partial molar quantities for the {\em real atoms}.  Even though vacancies are zero-mass, so they do not appear explicitly in the  partition function integral
(\ref{Definition2Z}) expression, this does not prevent them from showing up in the final parametric outcome $F(N_0; N_1, ..., N_C)$. $F(N_0; N_1, ..., N_C)$ is clearly a homogeneous function of degree 1, 
\begin{equation}
    F(\lambda N_0; \lambda N_1, ..., \lambda N_C) = \lambda F(N_0; N_1, ..., N_C),
\end{equation}
so all the Gibbsian thermodynamics formalism apply to $N_0$ (the point defect) equally as to the other real atoms.

At zero stress/pressure, the {\em equilibrium} vacancy concentration is defined by
\begin{equation}
    0 \equiv \left. \frac{\partial G}{\partial N_0} \right|_{\{N_{c\neq 0}\},T,P=0} = \mu_0 = h_0 - T s_0.
\end{equation}
Generally, we know $s_0$ should consist of translational, chemical configurational, and vibrational contributions:
\begin{equation}
    s_0 \equiv s_0^{\rm trans} + s_0^{\rm chem} + s_0^{\rm vib},
    \label{VacancyEntropy}
\end{equation}
where the translational part must be
\begin{equation}
    s_0^{\rm trans} = -k_{\rm B}\ln X_0
\end{equation}
in the dilute limit, and in the same limit $s_0^{\rm chem}$, $s_0^{\rm vib}$ have finite limiting values:
\begin{equation}
   \lim_{X_0\rightarrow 0} s_0^{\rm chem} = s_{\rm 0,dilute}^{\rm chem}, \;\;
   \lim_{X_0\rightarrow 0} s_0^{\rm vib} = s_{\rm 0,dilute}^{\rm vib},
\end{equation}
in contrast to $s_0^{\rm trans}$ hosting a log-singularity in the same limit.

In the conceptual development so far, $\tilde{N}=N_{\rm site}\uparrow$ when $N_0\uparrow$: one is generally thinking about the large-number limit, i.e. large number of vacancies and even larger number of real atoms.  But for numerical calculations with atomistic simulations, periodic boundary condition (PBC) with relatively small supercell is clearly preferred.  In this case, $N_{\rm site}$ is fixed, and $N_0\uparrow$ needs to come at the cost of some $N_c\downarrow$.  Because we are using the fundamental expression (\ref{MicroscopicFreeEnergy}) that uses $\{\mu_c\}$,
with grand-canonical interpretation, it is not difficult to do $N_0\uparrow, N_c\downarrow$ exchange for such PBC small-supercell calculations. Also note that in (\ref{ConfigurationalVibrational}),(\ref{MicroscopicFreeEnergy}), 
$s^N$ is now a configuration with $N$ real atoms, distributed on $\tilde{N}$ sites.

The key now is an MC sampling algorithm.  Consider the following idea experiment:  we have a $\tilde{N}=M^3\tilde{n}$ sites system, where $M$ is a large number, and $\tilde{n}$ is the size of the smaller cell. But although $\tilde{n}$ is smaller, this cell is still large enough to contain $2\times$ cutoff distance of the chemical short-range order of all atoms, i.e., the extent of the chemical PDF (pair distribution function), beyond which distance there is practically no more chemical correlations. (If this is not true, one should seriously consider increasing the $\tilde{n}$ cell size).  Then, it is a good approximation to have all $M\times M\times M$ cells {\em initially} being exact copies of one $\tilde{n}$ cell.  This is the reason one can estimate the free energy of a large system by doing a small PBC calculation.  In reality, different $\tilde{n}$ cells will be in different chemical configurations
\begin{equation}
  s^{\tilde{N}} \equiv  s^{\tilde{n}} \otimes s^{\tilde{n}} \otimes ... \otimes s^{\tilde{n}}
\end{equation}
where different $s^{\tilde{n}}$s are different. But the point is that due to the weak correlation beyond a certain distance, these different $s^{\tilde{n}}$s can be factored out, as if they were independent.

If we perform GCMC in our mind on the $\tilde{N}$ system, we would be likely comparing $s^{\tilde{N}}(N_0)$ configuration with an adjacent $s^{\tilde{N}}(N_0+1)$ configuration, where there are $N_0 \ll M^3$ vacancies already distributed in the former configuration, and in the latter configuration there is one real atom (say type-$c$) in a particular $\tilde{n}$ cell replaced by an additional vacancy.  Because a vacancy defect is quite expensive enthalpically, it is unlikely that this particular $\tilde{n}$ cell contains another vacancy, since a monovacancy is already a rare event, and divacancies (V$_2$), trivacancies (V$_3$), etc. are even rarer.  In other words, in this $\tilde{n}$ cell we go from a fully dense
\begin{equation}
  s^{\tilde{n}}(n=\tilde{n}) \rightarrow s^{\tilde{n}}(n=\tilde{n}-1)
  \label{VacancyAdd}
\end{equation}
and this causes microscopic free-energy difference 
\begin{align}
  & \Delta f = k_{\rm B}T\ln\left( \int \frac{d(\delta q^n)}{\lambda_{Tc}^{3n_1}.. \lambda_{Tc}^{3n_c}.. \lambda_{Tc}^{3n_C}} 
    e^{-\frac{U(s^{\tilde{n}}(n=\tilde{n}) + \delta q^n)}{k_{\rm B}T}} \right)  \label{VacancyCreationFreeEnergyDifference} \\
 & -   k_{\rm B}T\ln\left( \int \frac{d(\delta q^n)}{\lambda_{Tc}^{3n_1}.. \lambda_{Tc}^{3(n_c-1)}.. \lambda_{Tc}^{3n_C}} 
    e^{-\frac{U(s^{\tilde{n}}(n=\tilde{n}-1) + \delta q^n)}{k_{\rm B}T}} \right) + \mu_c 
    \nonumber
\end{align}
We may then accept or reject this move (\ref{VacancyAdd}) based on $e^{-\Delta f/k_{\rm B}T}$.  Even if $\Delta f$ turns out to be 2 eV in (\ref{VacancyCreationFreeEnergyDifference}), for example, there is still a small probability to accept this move.  

$\Delta f$ has spectrality, meaning it tends to have a distribution, depending on 
which site to remove the atom, and what $s^{\tilde{n}}(n=\tilde{n})$ is to begin with.  In the simplest case of $C=1$ elemental crystal, however, the distribution of $\Delta f$ would be a delta-function $\delta(\Delta f - f_{\rm V})$, where $f_{\rm V}=f_{\rm V}(T)$ is the vacancy formation energy in a monatomic crystal. Let us review precisely how Metropolis MC works in this case, in an idea experiment.  There are $N_0 \ll M^3$ vacated cells, and $M^3-N_0$ dense cells.  One should go over all $M^3 \tilde{n}$ sites {\em equally}, including the already vacated $N_0$ sites - even though there is no atom on the site.  For a vacated site, we will ask whether to back-fill with an atom from the reservoir.  Given most typically $\Delta f >0$, this back-filling move $-\Delta f <0$ should reduce the energy, and so will be accepted by Metropolis MC with probability $1$.  On the other hand, a great majority of the filled sites will be accepted by Metropolis MC procedure with fixed probability $e^{-\frac{f_{\rm V}}{k_{\rm B}T}}$, since most of the attempt newly vacated sites will be well separated from the already-vacated sites.  Thus, when we run the Metropolis MC procedure, we will reach steady-state porosity when
\begin{equation}
 N_0 \times 1 \approx  (M^3 \tilde{n} - N_0) \times e^{-\frac{f_{\rm V}}{k_{\rm B}T}},
\end{equation}
where the $\approx$ comes from ignoring the small chance of closely adjacent vacancies.  Then we derived the well-known thermal equilibrium vacancy fraction 
\begin{equation}
   X_{\rm V}^{\rm eq} \equiv \frac{N_0^{\rm eq}}{M^3 \tilde{n}} \approx e^{-\frac{f_{\rm V}}{k_{\rm B}T}}
\end{equation}
for monatomic crystal.  The sanity check above is just to say that Metropolis MC moves must be {\em site-based} equally, regardless of whether there is even an atom on the site.  If we choose to sample sites {\em unequally}, for example undersample those already-vacated sites, or have some spatial preference for filled sites, then the steady-state result would not be correct. 

Note that Metropolis MC aims to sample thermodynamic equilibrium, and one does not need to be kinetically faithful, in the sense that one does not need to pay attention to transition-state theory (TST) formulas that express the {\em real physical rate} (in unit of ${\rm s}^{-1}$) of transitions.  We know that within harmonic TST (HTST)\cite{Li07}, any rate for the physical process of back-filling of vacancy would require computing the vibrational frequencies {\em in the presence} of a vacancy.  One does not need to do so however in GCMC sampling of the thermodynamic equilibrium, since exchanging atom with a reservoir is not even a physical process, and one does not need to care about real vacancy sink/sources such as surfaces, dislocations, grain boundaries, etc.  One just need to pay attention that in statistical mechanics of Boltzmann, {\em all sites in a single-sublattice crystal must have equal chance} in the Metropolis MC procedure.  This basic property must hold in $C>1$ single-sublattice crystal, since different atomic types and associated CSRO are just different ways to decorate a site, in ways that are not fundamentally different from 1/0 decoration of a monatomic crystal. (Another way to say the same thing is that monatomic crystal with porosity is already a binary alloy anyway).

With this understanding, we can work out the expression for $X_{\rm V}^{\rm eq}(T)$ of $C>1$ alloy, by running the idea experiment in the large $M^3\tilde{n}$ sites system.  There are two families of moves: exchange within a 
$\tilde{n}$ cell that preserves $(n_0; n_1, ..., n_C)$, and addition/backfill of vacancy in a $\tilde{n}$ cell.  It can be easily shown that combination of these two families generate a third class of moves: exchange of atoms between two $\tilde{n}$ cells.  The two families of moves can have different frequency preference in this idea experiment.  But within each family, fairness needs to be respected: that is, a move and reverse move must have equal chance when energetics are not considered.  Also, different spatial sites must been treated equally.

We can first run plenty of exchange moves, to achieve the right CSRO distribution (thermodynamic mean and thermodynamic fluctuations).  This part is not controversial, and equilibrium between the real atoms in a fully dense solid will be approached.  We then run vacancy addition/backfill moves, equally among all $M^3 \tilde{n}$ sites.  For the vacancy addition moves, it is clear that for all $\Delta f > 0$ in (\ref{VacancyCreationFreeEnergyDifference}), we will have a low chance of acceptance, but can still succeed occasionally with probability
\begin{equation}
   P_{\rm V} \equiv \left\langle \frac{\sum_{i=1}^{\tilde{n}} e^{-\frac{\Delta f({\rm site}\; i)}{k_{\rm B}T}}}{\tilde{n}} \right\rangle
\end{equation}
when we run over all (initially filled) sites, which will create $X_{\rm V} = P_{\rm V}$ porosities (they will be no backfill, because initially no vacant sites) after the first sweep of all sites.  But in the 2nd sweep of sites for vacancy addition/backfill moves, we will create additional vacancies $\approx M^3 \tilde{n} P_{\rm V}$ again in the idea experiment, but the first batch of vacancies will then be {\em all backfilled} because presumably all $\Delta f = -(\Delta f)_{\rm orig} < 0$ and all the backfilling moves will be accepted in the Metropolis algorithm.  In other words, even though vacancy creation 
$\Delta f$ in (\ref{VacancyCreationFreeEnergyDifference}) has spectrality, as long as they are all positive valued, no vacancy can survive the next round of Metropolis batch sweep. Thus, in this Metropolis MC gedankenexperiment, the vacancy batches just appear and disappear with precisely life of 1 sweep, replaced by the next batch of vacancies with basically no relations with the previous vacancies (since divacancies are known to be very rare).   We thus can see in this limit of all $\Delta f = -(\Delta f)_{\rm orig} < 0$, the thermal equilibrium vacancy concentration is simply
\begin{equation}
   X_{\rm V}^{\rm eq}(T) \approx \left\langle \frac{\sum_{i=1}^{\tilde{n}} e^{-\frac{\Delta f({\rm site}\; i)}{k_{\rm B}T}}}{\tilde{n}} \right\rangle,
   \label{ThermalEquilibriumMonovacancy}
\end{equation}
based on the Metropolis MC gedankenexperiment.  The $\left\langle \right\rangle$ averaging in (\ref{ThermalEquilibriumMonovacancy}) must be based on a PBC ensemble of plenty of prior atom-preserving exchanges in the $\tilde{n}$-site supercell, i.e. sampling of a well-equilibrated, fully dense crystal, where CSRO between the real atoms are fully developed.  The definition of this PBC ensemble $\left\langle \right\rangle$  is non-controversial because it is where ab initio CALPHAD (Computer Coupling of Phase Diagrams and Thermochemistry) type approaches are based on, and also how the chemical potential $\{\mu_c(T)\}, c=1..C$ can be derived without consideration of vacancies.

In the Metropolis MC gedankenexperiment, one should of course repeat the (a) exchange, and (b) vacancy addition/backfill sweeps, multiple times, until convergence is reached.  But really it should not make a difference, because if the (a) sweeps are fully equilibrated, then $s^{\tilde{n}}(n=\tilde{n})$ satisfy Boltzmann distribution, and then it can be shown that the (b) sweep, even in the first sweep, generates Boltzmann distribution immediately, because free-energy changes are additive.  Then future sweeps, including atom-vacancy exchanges, would not make a difference either, statistically speaking, since energy changes are still additive (``Once Boltzmann, always Boltzmann", i.e. a trap distribution, which is a feature of equilibrium).  Another point is that in real Metropolis MC simulations, we do not do deterministic sweeps, but randomly pick lattice sites with equal probability (uniform lattice random). The deterministic sweep is a simplifying device.  But we can still track ``sweeps" on a site by looking at how many attempts a site has undergone. Even though the synchrony of the explanations above is gone, as we will reach equilibrium already in (a)+(b) sweep 1, none of these details would matter. The fact is that we have generated thermal equilibrium vacancies in the $M^3 \tilde{n}$ gedanken-cell.

\section{Results}

\subsection{General computational framework}\label{general_framework}  

Eq. (\ref{ThermalEquilibriumMonovacancy}) is conceptually simple and appealing. $\Delta f({\rm site}\; i)$ can be regarded as the {\em site-specific vacancy formation free energy}, whose value is explicitly expressed in the vibrational integrals Eq. (\ref{VacancyCreationFreeEnergyDifference}) as a free-energy difference. The thermal equilibrium vacancies can simply be regarded as having the Boltzmann distribution 
$e^{-\Delta f({\rm site}\; i)/k_{\rm B}T}$ on top of the fully dense $s^{\tilde{n}}(n=\tilde{n})$ configuration, based on these site-specific vacancy formation free energies.  The language of defect physics is modification of a reference state, and this is apparent in Eq. (\ref{ThermalEquilibriumMonovacancy}).

However, an analytically well defined and insightful expression does not necessarily mean it is easy to handle in numerical simulations.  The spectrality of $\Delta f({\rm site}\; i)$ in a Boltzmann distribution means extreme-value statistics could be important, i.e., those with the smallest $\Delta f({\rm site}\; i)$ for a specific $s^{\tilde{n}}(n=\tilde{n})$ configuration could totally dominate over all others sharing the same $s^{\tilde{n}}(n=\tilde{n})$ origin, in (\ref{VacancyAdd}). Furthermore, different $s^{\tilde{n}}(n=\tilde{n})$ origins are not equally distributed to begin with.  To analyze the net distribution of vacated configurations resulting from Eq. (\ref{ThermalEquilibriumMonovacancy}), let us write it out explicitly
\begin{equation}
   {\rm Prob}(s^{\tilde{n}}(n=\tilde{n}-1)) \propto \left\langle e^{-\frac{\Delta f({\rm site}\; i)}{k_{\rm B}T}} \right\rangle_{s^{\tilde{n}}(n=\tilde{n})},
   \label{ThermalEquilibriumMonovacancy2}
\end{equation}
which, given the definition of $\Delta f({\rm site}\; i)$ as microscopic free-energy difference, is simply
\begin{equation}
   {\rm Prob}(s^{\tilde{n}}(n=\tilde{n}-1)) \propto e^{-\frac{f({\rm site}\; i) - \sum_{c=1}^C n_c \mu_c }{k_{\rm B}T}}.
   \label{ThermalEquilibriumMonovacancy3}
\end{equation}
Eq. (\ref{ThermalEquilibriumMonovacancy3}) is very assuring.  It says the distribution of different vacated configurations is just according to its total free energy, and is independent of the {\em reference} or the
{\em path} of creating it.

There is indeed an issue of {\em path multiplicity} in creating a particular $s^{\tilde{n}}(n=\tilde{n}-1)$ configuration.  For example in 
CrCoNi alloy, $C=3$, one can create the same $s^{\tilde{n}}(n=\tilde{n}-1)$ from three different reference configurations, or $s^{\tilde{n}}(n=\tilde{n})$'s, where the vacated site is occupied by Cr, or occupied by Co, or occupied by Ni.  These three origins will have different free energies, as well as different (but compensating) site-specific vacancy formation free energies, but all three paths will result in the same final configuration.  It is true that going from Eq. (\ref{ThermalEquilibriumMonovacancy2}) to Eq. (\ref{ThermalEquilibriumMonovacancy3}) requires summing over all three 3 paths' contributions.  But writing things out shows that all three contributions have the same value, in other words a simple factor of $3\times$ is involved.  And this $3\times$ is true for all 
$s^{\tilde{n}}(n=\tilde{n}-1)$'s in the final expression of  Eq. (\ref{ThermalEquilibriumMonovacancy3}): given we use $\propto$ in Eq. (\ref{ThermalEquilibriumMonovacancy3}), such constant factor of $3\times$ would not matter. 

Eq. (\ref{ThermalEquilibriumMonovacancy3}), which can be named the {\em reference-free} version of vacancy configurational distribution, may be used to appreciate the extreme-value nature of the vacancy configurational dispersity (the $s_0^{\rm chem}$ term in (\ref{VacancyEntropy})).  The extreme 
end (support) of the vacancy distribution is
\begin{align}
   & \arg\min [ f({\rm site}\; i) - \sum_{c=1}^C n_c \mu_c ]
   \nonumber\\
= & \arg\min \left [ -   k_{\rm B}T\ln\left( \int \frac{d(\delta q^n)}{\lambda_{Tc}^{3n_1}.. \lambda_{Tc}^{3(n_c-1)}.. \lambda_{Tc}^{3n_C}} 
     \right.\right. \nonumber\\ 
    & \left. \left. e^{-\frac{U(s^{\tilde{n}}(n=\tilde{n}-1) + \delta q^n)}{k_{\rm B}T}}\right) - \sum_{c=1}^C n_c \mu_c \right]
 \label{VacancyMin}
\end{align}
and a {\em particular} vacated geometry ${\rm V}_{\rm min}$ and its symmetry-related variants (degeneracy factor $d_{{\rm V}_{\rm min}}$) will likely win this contest rigorously, for a particular $T$.  It would be interesting to render ${\rm V}_{\rm min}$ with ``core-shell like" chemical decorations, as well as its degeneracy factor $d_{{\rm V}_{\rm min}}(T)$, as a function of the temperature.  We note the following features regarding the spectrality and spectrality support: 
\begin{enumerate}
    
    \item By the definition Eq. (\ref{VacancyMin}), there can be discontinuous transitions in ${\rm V}_{\rm min}$ and $d_{{\rm V}_{\rm min}}$ as a function of $T$, similar to first-order phase transitions.

    \item Whether $({\rm V}_{\rm min}, d_{{\rm V}_{\rm min}})$ species dominate over some $({\rm V}_{\rm compete}, d_{{\rm V}_{\rm compete}})$ species in the population depends on the energy gap, as well as degeneracy factors.  If the energy gap is  larger than $k_{\rm B}T\ln(d_{{\rm V}_{\rm compete}}/d_{{\rm V}_{\rm min}})$, then ${\rm V}_{\rm min}$ will have larger population, and vice versa.

    \item ${\rm V}_{\rm min}$ and ${\rm V}_{\rm compete}$ will tend to have different $(n_1, ..., n_C)$'s, i.e. they are different in total chemical composition locally, i.e. possessing different Gibbsian segregation, as well as different CSROs (segregation patterns).
    
    \item One can also group many $({\rm V}_{\rm compete}, d_{{\rm V}_{\rm compete}})$'s within a certain absolute energy intervals, to compete with $({\rm V}_{\rm min}, d_{{\rm V}_{\rm min}})$. Depending on the energy gap and summed degeneracies, $({\rm V}_{\rm min}, d_{{\rm V}_{\rm min}})$ may still win in thermal equilibrium population compared to this group of competitors.  The situation is similar to molecular dispersity in petrochemical cracking.  Sometime a particular small molecule like ethylene ${\rm C}_2{\rm H}_4$ can win, occupying more than 50wt\% of the products; in other situations a large group of highly miscellaneous molecules can win. 

\end{enumerate}

In any case, solving (\ref{VacancyMin}) is a combinatorial optimization problem, and may not be trivial, in the sense that one can miss the global minimum numerically. This may be particularly catastrophic at low temperatures, where missing the ``correctly dressed" vacancy configuration ${\rm V}_{\rm min}$ may lead to severe underestimation of $X_{\rm V}^{\rm eq}$ and associated kinetic relaxation rates.  Below, we will use equiatomic CrCoNi alloy in face-centered cubic structure to illustrate these issues.

\subsection{Monte Carlo Sampling}

\begin{figure}[!htb]
    \centering
    \includegraphics{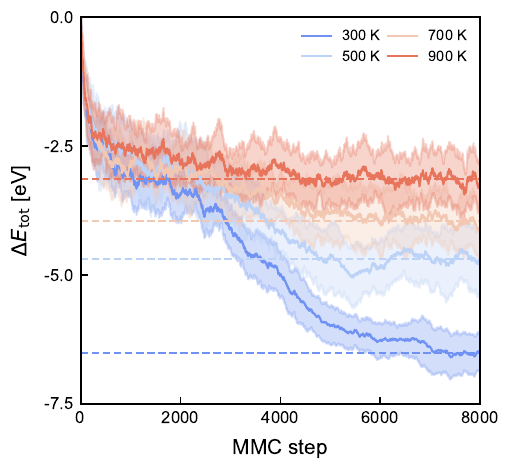}
    \caption{
    Monte Carlo sampling of low-energy configurations of CrCoNi. Energy minimization of CrCoNi at various temperature with a simulated annealing schedule, defined as $T = 1200 - \frac{1200 - T_{\rm anneal}}{4500} \times {\rm step}$ over 4500 simulation steps, followed by equilibration at a constant annealing temperature $T_{\rm anneal}$ for an additional 3500 steps. The dashed lines denote the energies obtained by averaging the last 1000 frames of each simulation.
}
    \label{fig:mmc}
\end{figure}

To numerically evaluate the equilibrium vacancy concentration expression Eq.~\eqref{ThermalEquilibriumMonovacancy}, we first sample the atomic structures of the $\tilde{n}$-site supercell (without vacancy) under thermal equilibrium. In the following calculations, we approximate the local free energy of each atomic configuration as the potential energy minimum, which means the entropy effects are not considered. The sampling is realized by the Metropolis Monte Carlo (MMC) simulation, as shown in Fig.~\ref{fig:mmc}. The atomic configurations in a supercell are initialized as fully random solid solution. As the number of sites in an 256-atom supercell is not divided by 3, we generate an ensemble with 5 Cr$_{85}$Co$_{85}$Ni$_{86}$, 5 Cr$_{85}$Co$_{86}$Ni$_{85}$ and 5 Cr$_{86}$Co$_{85}$Ni$_{85}$ at each temperature as initial atomic configurations to maintain exact equiatomic stoichiometry. All the initial configuration is then followed by a Monte Carlo simulation of 8,000 MMC steps.

In each simulation step, we randomly swap two atoms in a 256-atoms supercell with equal probability. The acceptance rate is set as $\min{(1, e^{-\Delta E/k_{\rm B}T} )}$, where $\Delta E$ is the change in potential energy after the swap. Specifically, we evaluate $\Delta E$ by the \texttt{MACE-MP-0}~\cite{batatia2023foundation} universal interatomic potential. Atomic configurations before and after the SWAP are both relaxed to the local energy minima with a force threshold of 0.03 eV/\AA, and $\Delta E$ is then evaluated as the potential energy difference before and after the atom swap. As we can see from Fig.~\ref{fig:mmc}, the potential energies of atomic systems converge to plateau values after 7000 MMC steps with a certain degree of random fluctuations. After the MMC simulation converges, the simulation trajectory samples the thermal equilibrium distribution of atomic configurations (without considering the vibrational entropy effect). Therefore, we take the last 1000 MMC steps (from 7000 to 8000 steps) to form our dataset to evaluate the average in Eq.~\eqref{ThermalEquilibriumMonovacancy}. The equilibrium configurations have strong short-range order (SRO), as shown in Fig.~S1-S3 in SI. The average SRO under equilibrium exhibits consistent magnitude and temperature dependence as the literature~\cite{cao2024capturing}, as shown in Fig. S4, validating that our MMC sampled ensemble is well converged.

\begin{figure}[!htb]
    \centering
    \includegraphics{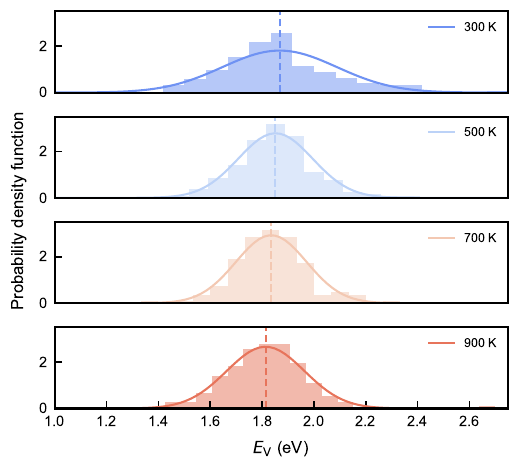}
    \caption{
    Distribution of vacancy formation energy ($E_{\rm V}$) with different temperatures. The solid lines represent Gaussian fits to the distributions, while the dashed lines indicate the corresponding mean values.
}
    \label{fig:sro_dos}
\end{figure}

After obtaining the atomic configurations, we evaluate $\Delta f({\rm site} \ i)$ in Eq.~\eqref{ThermalEquilibriumMonovacancy} of the sampled atomic configurations. Again, we approximate the free energy for vacancy formation by the vacancy formation energy $E_{\rm V}(\text{site}\ i)$, which is the energy difference before/after removing a site-$i$ atom, plus the chemical potential of the element at the site $i$. In order to get high-accuracy values, we implement the DFT calculations to evaluate the energies of the sampled configurations following the procedure below:

1. Randomly sample an atomic configuration $s$ with no vacancy from the last 1000 frames of the MMC trajectory.

2. Randomly remove an atom at the site $i$ in the supercell to form a vancancy-containing configuration $s'$.

3. Using DFT to relax both configuration $s$ and $s'$. See SI section III for the detailed DFT calculation parameters.

4. Obtain the converged DFT energies $E(s)$ and $E(s')$. The chemical potential of each element is evaluated with the substitution method, as detailed in SI section I. The expression $e^{-\frac{E_{\rm V}}{k_{\rm B}T}}$ is then evaluated for this data point with $E_{\rm V}\equiv E(s')-E(s)+\mu_{c(s\to s')}$.

5. Iterate through this process to collect data points and obtain the average in Eq.~\eqref{ThermalEquilibriumMonovacancy}.

Following this procedure, 720 DFT data points of $e^{-\frac{E_{\rm V}}{k_{\rm B}T}}$ are collected to sample its average. The sampled distribution of $E_{\rm V}$ approximately follow the Gaussian distribution, as shown in Fig.~\ref{fig:sro_dos}. The average value of the $E_{\rm V}$ distribution is negatively related to the temperature, with lower temperature exhibiting higher average vacancy formation energy. This is because stronger SRO forms at lower temperature, which decrease the average interaction energy between the nearest-neighbor atoms in the alloy. The decreased (more negative) interaction energy leads to a higher energy cost of removing an atom in the alloy (higher $E_{\rm V}$). Our predicted average $E_{\rm V}$ is well consistent with the previous literature~\cite{guan2020chemical}, validating our calculation procedure. 

\subsection{Equilibrium Vacancy Concentration}
\begin{figure}[!htb]
    \centering
    \includegraphics{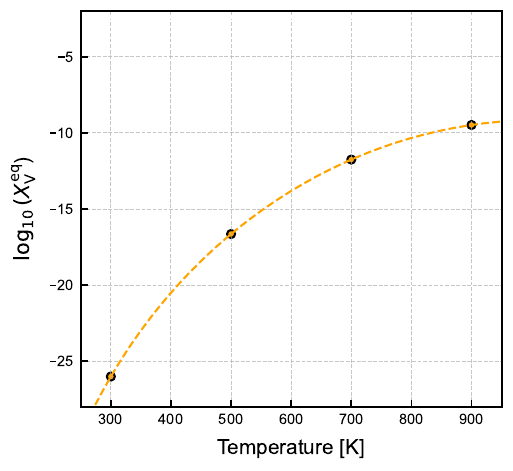}
\caption{
Change in equilibrium vacancy concentration $X_{\rm V}^{\rm eq}(T)$ with temperature.
}
    \label{fig:X_vac}
\end{figure}
The equilibrium vacancy concentration $X_{\rm V}^{\rm eq}(T)$ is then evaluated using Eq.~\eqref{ThermalEquilibriumMonovacancy} and the sampled data in Fig.~\ref{fig:sro_dos}. The equilibrium vacancy concentration $X_{\rm V}^{\rm eq}$ at 300 K, 500 K, 700 K,  and 900 K are evaluated as $9.9\times 10^{-27}$, $2.3\times 10^{-17}$, $1.8\times 10^{-12}$, $3.4\times 10^{-10}$, respectively, as shown in Fig.~\ref{fig:X_vac}. In experiments, the CrCoNi alloy is usually annealed at a temperature of about 600 - 900 K~\cite{li2023evolution}. As we can see, the equilibrium vacancy concentration at this regime is approximately in a range of $0.1-10$ ppb (percent per billion).

The temperature dependence of $\log{X_{\rm V}^{\rm eq}(T)}$ is interpolated by Lagrangian polynomial. The fitted results are shown below:
\begin{equation}
\begin{aligned}
    \log{X_{\rm V}^{\rm eq}(T)} &= aT^3 + bT^2 + cT + d, \\
    a &= 4.20\times 10^{-9}, \\
    b &= -7.83\times 10^{-6}, \\
    c &= 4.72\times 10^{-3}, \\
    d &= 7.24\times 10^{-1}.
    \end{aligned}
\end{equation}
The derived equilibrium vacancy concentration is used in our recent study for the short-range order kinetics in the CrCoNi alloy~\cite{chun2024learning}, which predicts the evolution timescale of the SRO in the alloy. Our prediction is well consistent with the experimental literature, providing validation to our method to calculate the equilibrium vacancy concentration.

\section{Discussion}

In this work, we developed a computational scheme to evaluate the equilibrium vacancy concentration in multi-principle element alloys. The method is applied to the equiatomic CrCoNi alloy, which provides the equilibrium vacancy concentration from 300 to 900 K. Equations similar or equivalent to Eq.~\eqref{ThermalEquilibriumMonovacancy} are also derived and evaluated in recent papers~\cite{lee2025modeling,li2024vacancy} for other multi-principal element systems. The differences of our work are in two aspects. On the one hand, our statistical mechanics derivation can in principal consider the vibrational entropy effects, which is not included in the previous work. On the other hand, we utilize universal machine learning potential together with the DFT calculations to provide high-accuracy evaluation to the potential energy landscape, while the previous work used empirical methods such as the effective interaction model~\cite{li2022magnetochemical} or embedded cluster expansions~\cite{muller2025constructing} to evaluate the potential energy landscape. 

Although the theoretical derivation of Eq.~\eqref{ThermalEquilibriumMonovacancy} is rigorous, we neglect the entropy effect in our numerical simulations. This could lead to slight deviation of the derived equilibrium vacancy concentration, either from the vibrational or configurational entropy effects. Discussion on the rigorous method for free energy calculations is provided in SI section S2. However, in the considered temperature range in our numerical calculations, we estimate the entropy effect to be small compared to the vacancy formation energy. Therefore, we leave more thorough numerical study of the entropy effect to future work. 

The equilibrium vacancy concentration at elevated temperature of high-entropy alloy is important to the
diffusive relaxation kinetics in alloy processing, as higher vacancy concentration will lead to faster diffusive relaxation processes. In our recent work in connection with this paper, we studied the SRO formation kinetics of the CrCoNi alloy during annealing at different temperature~\cite{chun2024learning}, where we use the equilibrium vacancy concentration derived in this paper. With the vacancy concentration, we derived SRO formation timescale consistent with the previous experimental observation. This also validate the prediction accuracy of $X_{\rm V}^{\rm eq}(T)$ we derived in this work. The computational method described in this work is generally applicable to different multi-principal element alloys.

\section{Code Availability}
The codebase for running atomistic simulations is available at \href{https://github.com/learningmatter-mit/RLVacDiffSim}{https://github.com/learningmatter-mit/RLVacDiffSim}. 

\section{Acknowledgements}
J.L. and H.T. acknowledge support by NSF DMR-1923976. This material is based upon work supported by the Under Secretary of Defense for Research and Engineering under Air Force Contract No. FA8702-15-D-0001. Any opinions, findings, conclusions or recommendations expressed in this material are those of the author(s) and do not necessarily reflect the views of the Under Secretary of Defense for Research and Engineering. This research used resources of the National Energy Research Scientific Computing Center (NERSC), a Department of Energy Office of Science User Facility using NERSC award GenAI@NERSC and MIT SuperCloud cluster. H.T. acknowledges support from the Mathworks Engineering Fellowship. 

\bibliography{bibliography}

\clearpage
\pagebreak
\setcounter{section}{0}
\setcounter{equation}{0}
\setcounter{figure}{0}
\setcounter{table}{0}
\setcounter{page}{1}
\makeatletter
\renewcommand{\theequation}{S\arabic{equation}}
\renewcommand{\thesection}{S\arabic{section}}
\renewcommand{\thefigure}{S\arabic{figure}}

\section{Supplementary Information }%
\subsection{Chemical potential and vacancy formation energy}\label{sec:chem_pot}
The elemental chemical potentials $\mu_{\rm Cr}$, $\mu_{\rm Co}$, and $\mu_{\rm Ni}$ in CrCoNi are evaluated using the Widom-type substitution method~\cite{wang2023effect, luo2025determinants}. 
Given a reference configuration with total energy $E_{\rm ref}$ and atomic counts $N_{\rm Cr}$, $N_{\rm Co}$, and $N_{\rm Ni}$, we define substitution energies $E_{A \to B}$ as the energy change associated with transmuting one atom of type $A$ into type $B$.
The chemical potential differences are approximated by symmetric combinations of these substitution energies:
\begin{equation}
\mu_{\rm Cr} - \mu_{\rm Co} = \frac{1}{2} \left( E_{\rm Co \rightarrow Cr} - E_{\rm Cr \rightarrow Co} \right),
\end{equation}
and similarly for other pairs.

The absolute chemical potentials are determined by solving the following total energy constraint:
\begin{equation}
E_{\rm ref} = N_{\rm Cr} \mu_{\rm Cr} + N_{\rm Co} \mu_{\rm Co} + N_{\rm Ni} \mu_{\rm Ni}.
\end{equation}
Solving this linear system yields the following expressions:
\begin{equation}
\begin{aligned}
\mu_{\rm Cr} &= \frac{1}{N_{\rm tot}} \Bigg[ E_{\rm ref} 
- \frac{N_{\rm Co}}{2} \left( E_{\rm Co \rightarrow Cr} - E_{\rm Cr \rightarrow Co} \right) \\
&\hspace{2.2cm}
- \frac{N_{\rm Ni}}{2} \left( E_{\rm Ni \rightarrow Cr} - E_{\rm Cr \rightarrow Ni} \right) \Bigg], \\
\mu_{\rm Co} &= \frac{1}{N_{\rm tot}} \Bigg[ E_{\rm ref} 
- \frac{N_{\rm Cr}}{2} \left( E_{\rm Cr \rightarrow Co} - E_{\rm Co \rightarrow Cr} \right) \\
&\hspace{2.2cm}
- \frac{N_{\rm Ni}}{2} \left( E_{\rm Ni \rightarrow Co} - E_{\rm Co \rightarrow Ni} \right) \Bigg], \\
\mu_{\rm Ni} &= \frac{1}{N_{\rm tot}} \Bigg[ E_{\rm ref} 
- \frac{N_{\rm Cr}}{2} \left( E_{\rm Cr \rightarrow Ni} - E_{\rm Ni \rightarrow Cr} \right) \\
&\hspace{2.2cm}
- \frac{N_{\rm Co}}{2} \left( E_{\rm Co \rightarrow Ni} - E_{\rm Ni \rightarrow Co} \right) \Bigg]
\end{aligned}
\end{equation}
where $N_{\rm tot} = N_{\rm Cr} + N_{\rm Co} + N_{\rm Ni}$.

We calculated the substitution energies from equilibrium configurations sampled by the MMC calculations at 300, 500, 700, and 900 K, as shown in Fig.~1 in the main text. 360 configurations are randomly sampled from the last 1000 frames in the MMC trajectory for each temperature, and then DFT calculations are implemented to evaluate the energy change after a randomly sampled element substitution. The reference energy $E_{\rm ref}$ is evaluated directly by the average of the sampled energies without element substitution. The calculated chemical potential of each element at different temperature is shown in Table~\ref{table:chemical_potentials}).

\subsection{Rigorous methods for free energy calculations}

There are several thermodynamic integration (TI) methods for both
elemental solids and alloys. The critical step in all methods is to
find a suitable reference system for which the free energy is either
known or can be easily and accurately computed. The TI methods integrate
the free energy starting from the reference state.

The three most common reference states are: (1) a set of isolated
(Einstein's) harmonic oscillators, (2) a perfectly harmonic solid,
and (3) an ideal system of non-interacting particles. The most common
integration methods are: (1) Frenkel-Ladd method (aka $\lambda$-integration),
(2) Integration along the temperature axis using the Gibbs-Helmholtz
equation, (3) integration along a chemical composition path, and (4)
Gibbs-Duhem integration.

The classical TI for alloys consists of two steps. First, we calculate
the free energy of one of the alloy components as a function of temperature; see, for example, \cite{Mishin04a,Williams06,Purja-Pun-2012,Howell2012}.
One way to accomplish this is to start with a relatively low temperature
$T_{0}$ at which the harmonic or (quasi-harmonic) approximation can
be applied and compute the harmonic free energy $F_{\mathrm{harm}}$
at this temperature. Then we run MD or MC to compute the system
energy $E(T)$ as a function of the temperature above $T_{0}$. The pressure
is assumed to be zero for simplicity. The free energy as a function
of temperature is obtained by the Gibbs-Helmholtz integration:
\begin{equation}
F(T)=F_{\mathrm{harm}}\dfrac{T}{T_{0}}-T\int_{T_{0}}^{T}\dfrac{E(T^{\prime})}{T^{\prime2}}dT^{\prime}.\label{eq:42}
\end{equation}
In practice, $E(T)$ can be well approximated by a cubic function
\begin{equation}
E(T)=E_{0}+AT+BT^{2},\label{eq:43}
\end{equation}
which simplified Eq.(\ref{eq:42}) to
\begin{equation}
F(T)=F_{\mathrm{harm}}\dfrac{T}{T_{0}}+E_{0}\left(1-\dfrac{T}{T_{0}}\right)-B(T-T_{0})-AT\ln\left(\dfrac{T}{T_{0}}\right).\label{eq:44}
\end{equation}

At the next step, we integrate the free energy with respect to the
chemical composition at a fixed temperature \citep{Mishin04a,Williams06,Purja-Pun-2012,Howell2012}.
Taking, for example, component 1 as a reference, we have
\begin{equation}
F(T,x_{1},...,x_{K})=F(T)+\sum_{k=1}^{K}\int_{0}^{x_{k}}(\mu_{k}-\mu_{1})dx_{k}.\label{eq:45}
\end{equation}
To perform the integration, we need to know the chemical composition
$(x_{1},...,x_{K})$ as a function of the diffusion potentials $(\mu_{k}-\mu_{1})$,
or the diffusion potentials as a function of the chemical composition.
This function can be obtained by one of two methods. In the first method,
we run semi-grand canonical MC simulations in which we impose the
diffusion potentials and determine the chemical composition after
thermodynamic equilibrium is reached. The trial moves are random displacements
of atoms and attempts to change their chemical identity. For example,
if we attempt to switch the chemical species of a random atom from
1 to 2, the probability of the new microstate $s^{\prime}$ relative
to the old one $s$ is
\begin{equation}
\dfrac{P^{\prime}}{P}=\left(\dfrac{\lambda_{1}}{\lambda_{2}}\right)^{3}\exp\left(-\dfrac{u_{s^{\prime}}-u_{s}-(\mu_{2}-\mu_{1})}{k_{B}T}\right).\label{eq:46}
\end{equation}
The ratio $P^{\prime}/P$ is used to accept or reject the attempt
according to the Metropolis criterion. The prefactor $(\lambda_{1}/\lambda_{2})^{3}$
originates from the kinetic energy and is often ignored, creating
a systematic bias in the results.

In the second method, we first equilibrate the system by a canonical
MC simulation. The trial moves are atomic displacements and swapping
of atomic pairs. The relative probability of the initial and final
microstates is simply
\begin{equation}
\dfrac{P^{\prime}}{P}=\exp\left(-\dfrac{u_{s^{\prime}}-u_{s}}{k_{B}T}\right).\label{eq:47}
\end{equation}
Once equilibrium is reached, we continue the simulation and once in
a while switch the chemical species of a random atom. Such moves are
never accepted, but the exponential term below is averaged to give
a diffusion potential. For example, for the switches $2\rightarrow1$,
we obtain
\begin{equation}
\mu_{1}-\mu_{2}=k_{B}T\left(\dfrac{\lambda_{2}}{\lambda_{1}}\right)^{3}\ln\left\langle \exp\left(-\dfrac{u_{s^{\prime}}-u_{s}}{k_{B}T}\right)\right\rangle _{0},\label{eq:48}
\end{equation}
where the subscript 0 indicates that we run the simulations in the
ensemble of the initial state and never accept transitions to other
states with switches chemical identities. 

Whichever method is used, the relation between the diffusion potentials
and composition along a chosen path is represented by a set of 20-30
points and used for a numerical calculation of the integral in Eq.(\ref{eq:45}).
To cross-check the results, the calculation is repeated for several
different paths leading to the same chemical composition. In addition,
the procedure can be repeated starting from a different end member
to check that it leads to nearly the same free energy. This method
has its pros and cons. It is conceptually straightforward, but requires
extensive calculations. More importantly, it only works if there is
a continuous path from the chemical composition we are interested
in to one of the end members without a phase transformation in between.

The Gibbs-Helmholtz integration in Eqs.(\ref{eq:42})-(\ref{eq:44})
can be performed for a fixed atomic configuration in an alloy. In
this case, the calculated free energy represents the atomic vibrations
(generally, anharmonic), but excludes the configurational entropy.

Returning to an elemental crystalline solid, an alternative approach
is to perform $\lambda$-integration using a set of identical harmonic
oscillators as the reference system. The free energy $F_{\mathrm{harm}}$
of such oscillators can be exactly calculated. The free energy of the
actual solid is obtained from the integral
\begin{equation}
F=F_{\mathrm{harm}}+\int_{0}^{1}\left\langle u-u_{\mathrm{harm}}\right\rangle _{\lambda}d\lambda.\label{eq:49}
\end{equation}
Here, $u$ is the potential energy of the actual solid and $u_{\mathrm{harm}}$
is the potential energy of the oscillators for the \emph{same} set of atomic positions.
The subscript $\lambda$ indicates that the average in angular
brackets is computed by running an MC simulation in the ensemble in
which the energy of the atomic configurations is given by $(1-\lambda)u+\lambda u_{\mathrm{harm}}$.
The oscillators are fixed at the average atomic positions $\mathbf{r}_{l}^{0}$
in the actual solid. The oscillator energy is given by $c(\mathbf{r}_{l}-\mathbf{r}_{l}^{0})^{2}$,
where $c$ is an elastic constant. The latter is adjusted to make
the reference state as close to the actual state as possible. 

The $\lambda$-integration method requires some care, especially with
regard to the center of mass of the system, which is well documented
in the literature. However, in general, the method is robust and widely
used. A generalization of this method is to replace the set of Einstein
oscillators with a single harmonic solid \citep{Zhou:2022aa,Jung:2023aa}.
The reference free energy can again be calculated analytically. 

It should be mentioned that, in some cases, the $\lambda$-integration
can be replaced by a simpler calculation using the free energy perturbation
(FPT) theory. In this theory \citep{Zwanzig:1954aa}, 
\begin{equation}
F=F_{\mathrm{harm}}-k_{B}T\ln\left\langle \exp\left(-\dfrac{u-u_{\mathrm{harm}}}{k_{B}T}\right)\right\rangle _{\mathrm{harm}}.\label{eq:50}
\end{equation}
The meaning of the square brackets is the same as above. The simulations
are run in the harmonic ensemble with occasional switches to the real
energy $u$. Such switches are recorded but never accepted. This is
a ``one-shot'' integration, which only works if the two states are
close to each other and their distributions significantly overlap.

Extension of the $\lambda$-integration method to alloys is very complex.
So far, the method has only been applied to elemental solids, as well
as alloys with a \emph{fixed} atomic configuration. The method only
captures the vibrational free energy and is silent on the configurational
entropy. In principle, one can apply this method to compute the (vibrational)
free energies of several (or many) atomic configurations and average
the vibration free energy. However, the configurational free energy
still remains unaccounted for. 

We can explore two possible extensions of the $\lambda$-integration
method to alloy. One approach is to create a new artificial (fictitious)
alloy composed of the same chemical components but forming a perfect
solution (zero heat of mixing). The configurational entropy of this
artificial alloy is ideal and is trivially known. Then we can proceed
in two steps. First, compute the (anharmonic) vibrational free energy
of the artificial alloy by $\lambda$-integration starting from a
harmonic state at a lower temperature. This was discussed above. By
adding an ideal configurational entropy term, we obtain the full free
energy $F_{0}$ of this system. This system can then be used as a reference
state. In the second step, we run canonical MC simulations for the
artificial alloy and occasionally switch the energy to that of the
actual alloy with nonzero heat of mixing and non-ideal configurational
entropy. The free energy of the actual alloy is the found from the
perturbation equation 
\begin{equation}
F=F_{\mathrm{0}}-k_{B}T\ln\left\langle \exp\left(-\dfrac{u-u_{\mathrm{0}}}{k_{B}T}\right)\right\rangle _{\mathrm{0}}.\label{eq:51}
\end{equation}
Of course, we can also run the simulation in the ensemble of the real
alloy and switch to the artificial one. If the overlap of distributions
is not strong enough, we can apply the $\lambda$-integration using
the same reference state:
\begin{equation}
F=F_{\mathrm{0}}+\int_{0}^{1}\left\langle u-u_{\mathrm{0}}\right\rangle _{\lambda}d\lambda,\label{eq:52}
\end{equation}
where the average in square brackets is calculated in the ensemble
with the energy $(1-\lambda)u+\lambda u_{\mathrm{0}}$. In either
case, the energy and its dispersion in the artificial alloy must be
adjusted to give the same average energy and about the same dispersion
as for the actual alloy. This will ensure rapid convergence in Eq.(\ref{eq:52})
or applicability of the perturbation formula (\ref{eq:51}). The caveat
is that the solid solution with the fictitious energy $(1-\lambda)u+\lambda u_{\mathrm{0}}$
can become unstable mechanically, or against a structural transformation
(e.g., FCC-HCP), or against separation into phases. It does not appear
that this approach was previously explored in the literature.

Another possibility is to create an ideal mixture of Einstein oscillators
centered at average atomic positions. The oscillators are designed
to represent the alloy elements as closely as possible. They are characterized
by different masses and spring constants, which should be adjusted
to approximate the vibrational spectrum of the real solution as closely
as possible. Special care should be taken to fix the centers of mass
of the real solution and the oscillators. Since the oscillators do
not interact with each other, the free energy of the reference system
is known analytically. The free energy of the actual alloy can be
obtained by $\lambda$-integration using Eq.(\ref{eq:52}). As above,
the risk is that the system with the fictitious energy at intermediate
$\lambda$ values can lose stability. Alternatively,
the free energy can be obtained in one shot from the perturbation formula (\ref{eq:51}).
For a binary system, this approach was explored in~\citep{Ogando:2002aa}.

In principle, yet another approach is to heat the solid solution to
such a high temperature that the short-range order would be negligible
and the configuration entropy could be approximated by the ideal Ansatz.
For systems with a strong short-range order, this approach seems to
be impractical. 

In summary, there are three options for computing the absolute values
of the free energy for a multicomponent solid solution by thermodynamic
integration:
\begin{enumerate}
\item Gibbs-Helmholtz integration for one of the chemical elements of the
solution followed by chemical integration along a path leading to
the current chemical composition.
\item Two-step $\lambda$-integration using an artificial perfect solid
solution with an ideal configurational entropy as a reference system.
\item One-step $\lambda$-integration using a mixture of harmonic oscillators
as a reference system.
\end{enumerate}
Method 1 is well-established, while methods 2 and 3 require exploratory
research. Each method has its pros and cons. The common challenge
in all three methods is to prevent loss of stability of the system
along the integration path. Using the FPT approach partially mitigates
the instability problem, but it only works when the reference system
is sufficiently close to the real one. 

Once the free energy is known, one can recover the individual chemical
potentials if they are needed. This can be achieved by computing the
diffusion potentials from Eq.(\ref{eq:48}) and combining them with
the Gibbs equation $F=\sum_{k}\mu_{k}N_{k}$ (zero pressure) to obtain
the absolute values of the individual chemical potentials.

\begin{table*}[ht]
\centering
\caption{Elemental chemical potentials $\mu_{\rm elem}$ (in eV) computed under different temperature from 300 K to 900 K.}
\label{table:chemical_potentials}
\begin{tabular}{ccccc}
\hline
\hline
\ \ \ Element \ \ \  &\ \ \  $\mu^{\rm 300 K}_{\rm elem}$  (eV)\ \ \  & \ \ \ $\mu^{\rm 500 K}_{\rm elem}$ (eV)\ \ \  & \ \ \ $\mu^{\rm 700 K}_{\rm elem}$ (eV) \ \ \ & \ \ \ $\mu^{\rm 900 K}_{\rm elem}$ (eV)\ \ \  \\
\hline
\hline
Cr & $-9.29$ & $-9.30$ & $-9.35$ & $-9.32$\\
\hline
Co & $-7.08$ & $-6.98$ & $-6.98$ & $-6.95$\\
\hline
Ni & $-5.44$ & $-5.50$ & $-5.44$ & $-5.48$ \\
\hline
\hline
\end{tabular}
\end{table*}

\begin{figure*}[!htb]
    \centering
    \includegraphics{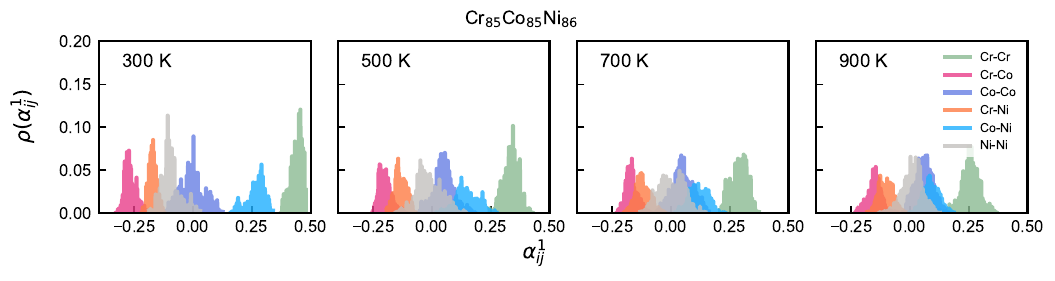}
    \caption{
    Distribution of WC parameters $\alpha_{ij}^1$ for $\rm Cr_{85}Co_{85}Ni_{86}$ at equilibrium across different temperatures.
}
    \label{fig:sro_dos_858586}
\end{figure*}

\begin{figure*}[!htb]
    \centering
    \includegraphics{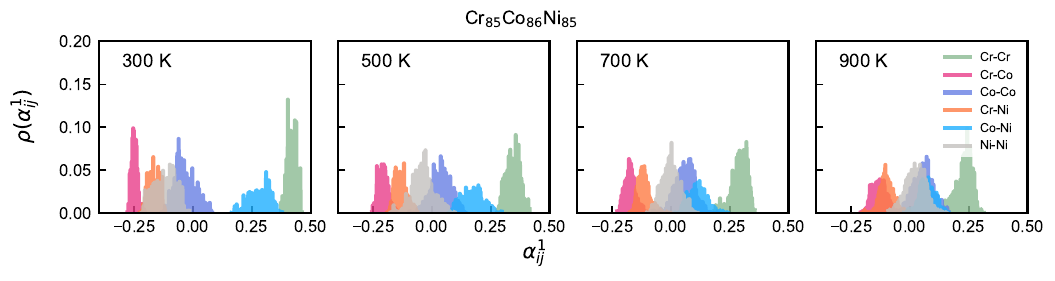}
    \caption{
    Distribution of WC parameters $\alpha_{ij}^1$ for $\rm Cr_{85}Co_{86}Ni_{85}$ at equilibrium across different temperatures.
}
    \label{fig:sro_dos_858685}
\end{figure*}
\begin{figure*}[!htb]
    \centering
    \includegraphics{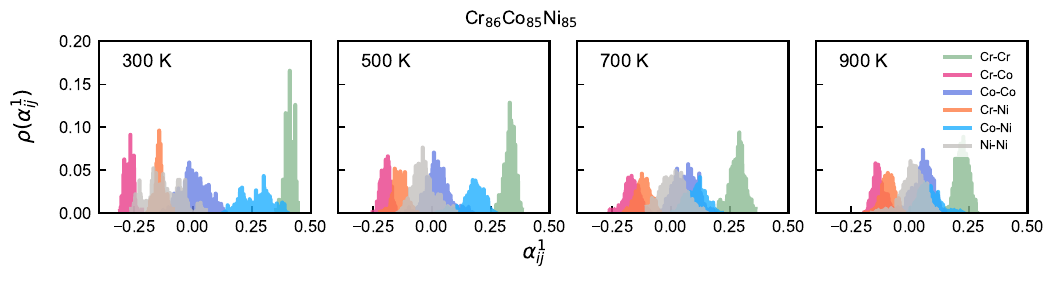}
    \caption{
    Distribution of WC parameters $\alpha_{ij}^1$ for $\rm Cr_{86}Co_{85}Ni_{85}$ at equilibrium across different temperatures.
}
    \label{fig:sro_dos_868585}
\end{figure*}

\begin{figure*}[!htb]
    \centering
    \includegraphics[width=0.5\textwidth]{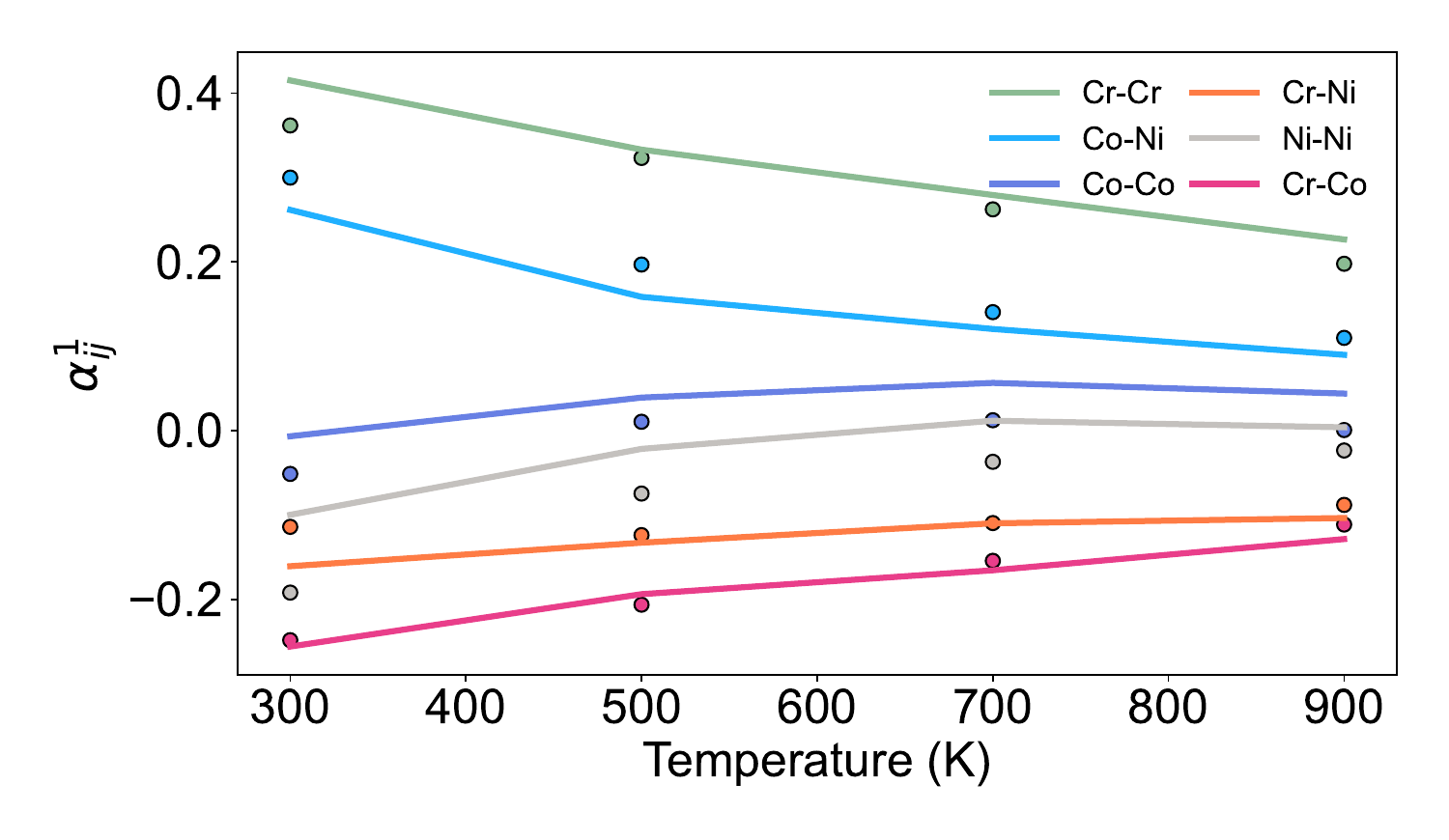}
    \caption{
Comparison of the WC parameter $\alpha_{ij}^{1}$. Solid lines represent the present results, showing average $\alpha_{ij}^{1}$ values at different temperatures, while the scatter points denote previously reported values~\cite{cao2025capturing}.
}
    \label{fig:sro_compare}
\end{figure*}

\end{document}